\documentclass[journal]{IEEEtran}

\usepackage{enumitem}
\usepackage[linesnumbered,ruled,lined]{algorithm2e}

\usepackage[
  colorlinks=true,
  linkcolor=blue,
  citecolor=blue,
  urlcolor=blue
]{hyperref}

\usepackage{booktabs,array,tabularx,makecell,threeparttable}
\newcolumntype{C}[1]{>{\centering\arraybackslash}m{#1}}
\newcolumntype{Y}{>{\centering\arraybackslash}X}

\AtBeginDocument{}

\usepackage{graphicx}% Include figure files
\usepackage{dcolumn}% Align table columns on decimal point
\usepackage{bm}% bold math

\usepackage[usenames,dvipsnames]{xcolor}

\usepackage{amsmath,amsfonts}
\usepackage{mathtools} % Include in preamble

\usepackage{array}
\usepackage[caption=false,font=normalsize,labelfont=sf,textfont=sf]{subfig}
\usepackage{textcomp}
\usepackage{stfloats}
\usepackage{url}
\usepackage{verbatim}
\usepackage{cite}
\usepackage{svg}
\usepackage[noend]{algpseudocode} 
\usepackage{etoolbox}
\usepackage{multirow}

\usepackage{float}
\usepackage{pgfplots}
\usetikzlibrary{shapes.multipart,intersections}
\usepackage{amssymb,amsthm,steinmetz}
\usepackage{mathrsfs}  
\usepackage{acronym}
\usepackage{upgreek,xspace}

\usepackage{tikz}
\usetikzlibrary{calc}
\makeatletter
\newcommand{\gettikzxy}[3]{%
  \tikz@scan@one@point\pgfutil@firstofone#1\relax
  \edef#2{\the\pgf@x}%
  \edef#3{\the\pgf@y}%
}
\makeatother

\usepackage[draft]{todonotes}

\usepackage{esvect}

\usepackage{times}
\usepackage{stmaryrd}
\usepackage{babel}
\usepackage{graphics}
\usepackage{gensymb}

\ifCLASSINFOpdf

\else

\fi

\usepackage[all=normal,paragraphs=tight,floats=normal,mathspacing=normal,wordspacing=tight,charwidths=tight,mathdisplays=normal,leading=normal]{savetrees}

\begin{document}

\title{Cross-Harmonic Ambiguity-Aligned Multiport Parameter Estimation for Time-Floquet RIS}

\author{Philipp~del~Hougne,~\IEEEmembership{Member,~IEEE}
\thanks{
P.~del~Hougne is with the Department of Electronics and Nanoengineering, Aalto University, 00076 Espoo, Finland, and also with Univ Rennes, CNRS, IETR - UMR 6164, F-35000, Rennes, France. (e-mail: philipp.del-hougne@univ-rennes.fr)
\textit{(Corresponding Author: Philipp~del~Hougne.)}}
\thanks{This work was supported in part by the Nokia Foundation (project 20260028), the ANR France 2030 program (project ANR-22-PEFT-0005), and the ANR PRCI program (project ANR-22-CE93-0010).}
}

% make the title area
\maketitle

\begin{abstract}
A time-Floquet reconfigurable intelligent surface (TF-RIS) periodically modulates its elements within a signaling interval, enabling frequency conversion and additional degrees of freedom compared with a conventional RIS. Time-Floquet multiport-network theory (TF-MNT) provides a physics-consistent model for TF-RISs that accounts for inter-element coupling, but its practical use requires estimating the underlying parameters when the TF-RIS design and radio environment are (partially) unknown. In this Letter, we propose a segmented estimation approach for constructing an accurate proxy TF-MNT model from end-to-end measurements. \textit{First}, with the TF-RIS operated as a conventional RIS, we estimate conventional proxy MNT parameters independently at each considered time-Floquet harmonic. \textit{Second}, under periodic time modulation, we align the inherent ambiguities among the per-harmonic conventional proxy MNT parameters, considering three measurement setups with different access to phase and harmonic information. Based on full-wave numerical simulations, we quantify the impact of the number of measurements and the noise level on the proxy-model accuracy. Finally, we demonstrate the performance loss incurred without the proposed ambiguity alignment in a canonical harmonic backscatter communications scenario.
\end{abstract}

\begin{IEEEkeywords}
Ambiguity, ambiguity-aware segmentation, harmonic beam steering, harmonic channel estimation, mutual coupling, multiport network theory, periodic time modulation, time-Floquet reconfigurable intelligent surface.
\end{IEEEkeywords}

\IEEEpeerreviewmaketitle

\section{Introduction}
\label{sec_introduction}

Reconfigurable intelligent surfaces (RISs) endow wireless channels with deterministic programmability. From an electromagnetic perspective, an RIS element is an antenna element whose feed port is terminated by a tunable load. This abstraction enables a physics-consistent description of RIS-parametrized channels in terms of multiport-network theory (MNT)~\cite{gradoni_EndtoEnd_2020}. A conventional RIS element's load is reconfigurable but kept fixed within a signaling interval. In contrast, a new generation of RISs, referred to as time-Floquet RISs (TF-RISs) in this Letter, is distinguished by the periodic modulation of its loads within a given signaling interval~\cite{zhang2018space,mizmizi2024wireless,verde2024rapidly,kuznetsov2025multifrequency,kuznetsov2026mutual}. This periodic load modulation introduces new degrees of freedom and frequency conversion; among the unique applications enabled by a TF-RIS, a simple example is harmonic beam steering~\cite{zhang2018space,kuznetsov2026mutual}. Recently, a physics-consistent description of TF-RIS-parametrized systems based on time-Floquet MNT (TF-MNT) was established~\cite{kuznetsov2025multifrequency,kuznetsov2026mutual}. 

While TF-MNT-based theoretical studies such as~\cite{kuznetsov2026mutual} can assume perfect knowledge of the TF-MNT parameters, in practice the acquisition of an accurate set of TF-MNT parameters for an experimentally given TF-RIS-based system will be a major concern. In the experimental reality, system details are unknown for one or multiple of the following reasons: (i) undisclosed proprietary TF-RIS designs; (ii) insufficiently documented characteristics of the tunable lumped elements; (iii) deviations from the expected design due to fabrication inaccuracies; (iv) partially or fully unknown geometry and material composition of the radio environment. Thus, the TF-MNT model parameters must be estimated experimentally. Usually, they cannot be measured directly because the loads are not connectorized and too numerous. Thus, an indirect experimental estimation of the TF-MNT parameters will be required.

For conventional RISs, significant efforts were recently dedicated to the indirect experimental estimation of an accurate set of proxy MNT parameters~\cite{sol2024experimentally,del2025experimental,del2026ambiguity,del2026reduced}. Importantly, these efforts go beyond conventional channel estimation because they additionally estimate the mutual coupling (MC) between the RIS elements and the characteristics of the available loads. As seen in the experiments in~\cite{rabault2024tacit,del2026reduced}, the MC can strongly depend on the radio environment such that even with a perfectly known RIS design the MC can remain unknown. The term ``proxy'' emphasizes inevitable ambiguities in the estimated parameters\footnote{To lift all ambiguities, one would require at least two known individual loads as well as the possibility to terminate pairs of neighboring RIS elements, as well as at least one RIS element and one accessible antenna, with known coupled loads~\cite{del2024virtual2p0,del2025wireless}.}; despite these ambiguities, the proxy parameters enable the accurate prediction of the physically measurable end-to-end channels for any admissible RIS configuration. Thus, these ambiguities are usually operationally irrelevant. However, the ambiguities require careful handling when the parameter estimation is segmented to ensure that the ambiguities of different segments are aligned~\cite{del2026ambiguity}. Moreover, ambiguity-sensitive bounds can be tightened by optimizing over the ambiguities [Sec.~IV.A,~\cite{salmi2026electromagnetically}]. Since existing techniques estimate the MNT parameters independently per frequency point, the ambiguities generally differ for each frequency point. For conventional RISs, this is usually not problematic because there is no frequency mixing; however, different per-frequency ambiguities could matter when seeking a compact state-space-based representation~\cite{deschrijver2008macromodeling} for a broadband channel parametrized by a conventional RIS. 

Meanwhile, since frequency mixing is a key feature of TF-RISs, mismatched cross-frequency ambiguities hinder the use of per-frequency proxy MNT parameters within the TF-MNT formalism. This motivates a cross-harmonic ambiguity alignment that we study in the present Letter. To accommodate constraints on peak computational complexity and memory usage, as well as to enable parallelization, we segment the TF-MNT parameter estimation. In a first step, operating the TF-RIS as a conventional RIS, we estimate proxy MNT parameters separately for each considered harmonic. In a second step, operating the TF-RIS with periodic time modulation, we perform a cross-harmonic ambiguity alignment. Besides a full cross-harmonic ambiguity alignment, we also consider partial cross-harmonic ambiguity alignments that are sufficient and experimentally less demanding when only the complex end-to-end channel at the fundamental frequency or only the intensities of the harmonic end-to-end channels matter.

Our contributions are summarized as follows.
\textit{First}, we present a technique for estimating an ambiguity-aligned set of TF-MNT parameters, fully accounting for practical hardware constraints: MC between TF-RIS elements, a finite number of dispersive load states, structural scattering of antennas and RIS elements, and environmental scattering. Moreover, we consider three experimental measurement modalities.
\textit{Second}, based on a full-wave numerical simulation, we systematically examine the achieved accuracy of harmonic end-to-end channel predictions with our proxy TF-MNT parameters as a function of the number of measurements and noise level.
\textit{Third}, we examine the performance loss in TF-RIS-aided communications without cross-harmonic ambiguity alignment.

\textit{Notation:}
$\mathrm{diag}(\mathbf{a})$ denotes the diagonal matrix whose diagonal entries are $\mathbf{a}$.
$\mathrm{blkdiag}(\mathbf{A}_1,\dots,\mathbf{A}_a)$ denotes the block-diagonal matrix constructed from the matrices $\mathbf{A}_1,\dots,\mathbf{A}_a$.
$\mathbf{A}_\mathcal{BC}$ denotes the block of the matrix $\mathbf{A}$ whose row [column] indices are in the set $\mathcal{B}$ [$\mathcal{C}$]. 
$|\mathcal{B}|$ denotes the cardinality of the set $\mathcal{B}$.
$\|\cdot\|_1$ denotes the entrywise $\ell_1$ norm.
$^\top$ denotes the transpose operation.
$\mathbf{I}_a$ is the $a\times a$ identity matrix.
$\mathbf{0}$ denotes a matrix of suitable size whose entries are all zero.
$\jmath$ denotes the imaginary unit.

\section{System Model}
\label{sec_system_model}

\begin{figure}
    \centering
    \includegraphics[width=\columnwidth]{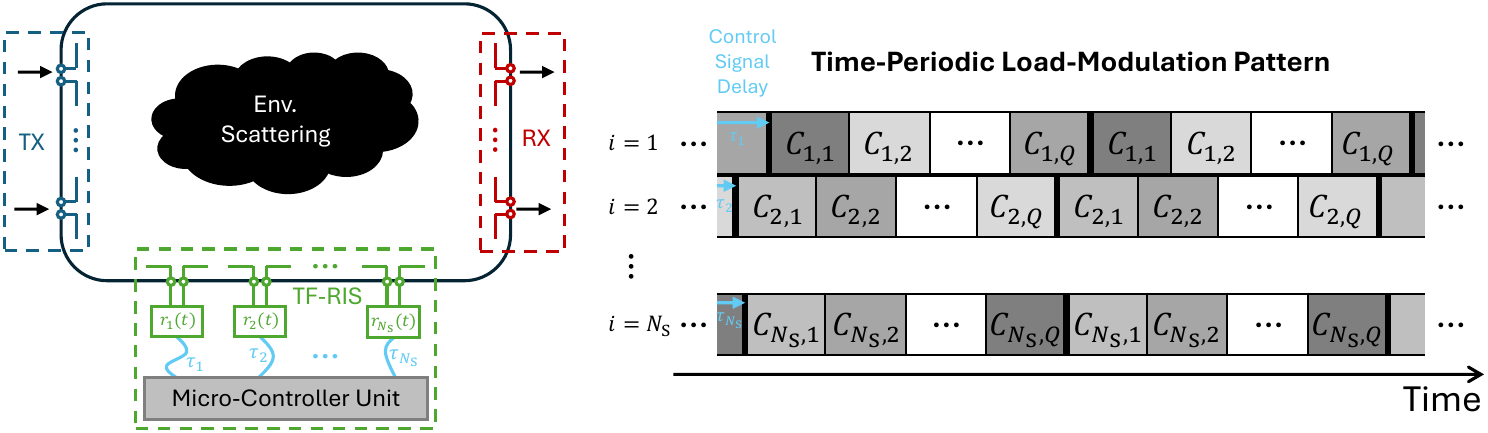}
    \caption{Left: MNT-based system model for a TF-RIS-parametrized radio environment. Right: Time-periodic load-modulation pattern.}
    \label{Fig1}
\end{figure}

In this section, we summarize based on~\cite{kuznetsov2025multifrequency,kuznetsov2026mutual} the TF-MNT model for a TF-RIS. We divide one modulation period \(T_\mathrm{m}\) into \(Q\) slots of duration \(\Delta=T_\mathrm{m}/Q\). We assume that each tunable element can realize \(P\) programmable states, and we specify the modulation pattern by \(\mathbf C\in\{1,\dots,P\}^{N_\mathrm{S}\times Q}\), where \(C_{i,q}\) is the state applied to the \(i\)th element during the \(q\)th slot. To accommodate imperfections of practical control circuits, we do \textit{not} assume that the slot boundaries are temporally aligned across all elements. Instead, we allow the control signal for the \(i\)th element to be delayed by \(\tau_i\), as illustrated in Fig.~\ref{Fig1}.

The periodic time modulation of the TF-RIS gives rise to frequency conversion, meaning that a signal at frequency $f_0$ can scatter into harmonic frequencies $f_h=f_0+h f_\mathrm{m}$, where $f_\mathrm{m}=1/T_\mathrm{m}$ is the modulation frequency of the TF-RIS and $f_\mathrm{m}\ll f_0$. In practice, we can only retain a finite set of harmonics $\mathcal{H}=\{h_1,\dots,h_{|\mathcal{H}|}\}\subset\mathbb{Z}$, which we choose symmetric around the fundamental harmonic $h=0$. 

The time-Floquet end-to-end channel matrix $\widetilde{\mathbf{H}}\in\mathbb{C}^{|\mathcal H|N_\mathrm{R}\times |\mathcal H|N_\mathrm{T}}$ characterizes the multi-frequency linear input-output relation of the TF-RIS-parametrized radio environment with $N_\mathrm{T}$ transmitters and $N_\mathrm{R}$ receivers: $\widetilde{\mathbf{b}}=\widetilde{\mathbf{H}}\,\widetilde{\mathbf{a}}$. Here, $\widetilde{\mathbf{a}}\in\mathbb{C}^{|\mathcal H| N_\mathrm{T}}$ and $\widetilde{\mathbf{b}}\in\mathbb{C}^{|\mathcal H| N_\mathrm{R}}$ are the multi-frequency input and output wavefronts, respectively. Specifically, $\widetilde{\mathbf{a}} = \left[{\mathbf{a}^{(h_1)}}^\top {\mathbf{a}^{(h_2)}}^\top \dots \ {{\mathbf{a}}^{(h_{|\mathcal H|})}}^\top \right]^\top \in \mathbb{C}^{|\mathcal H| N_\mathrm{T}}$, where $\mathbf{a}^{(h_i)}\in\mathbb{C}^{N_\mathrm{T}}$ is the input wavefront at frequency $f_{h_i}$, and $\widetilde{\mathbf{b}}$ is defined analogously. For a conventional RIS, $\widetilde{\mathbf{H}}$ would be block-diagonal due to the absence of frequency mixing; in contrast, for a TF-RIS, $\widetilde{\mathbf{H}}$ is generally \textit{not} block-diagonal.

To determine $\widetilde{\mathbf{H}}$, we begin by partitioning the TF-RIS-parametrized radio environment into three entities: (i) $N_\mathrm{A}=N_\mathrm{T}+N_\mathrm{R}$ antenna ports used to inject and/or receive waves, (ii) $N_\mathrm{S}$ tunable lumped elements, and (iii) the ensemble of all static scattering objects (including both environmental and structural scattering). Next, we describe each tunable lumped element as a ``virtual'' port terminated by a tunable load. Now, we can interpret the entire system as the connection of two multiport subsystems, one comprising all static components and one comprising all tunable components. We define all scattering parameters for a 50~$\Omega$ reference impedance.

Because the subsystem comprising all static components is linear and time-invariant, its time-Floquet scattering matrix is simply $\widetilde{\mathbf S}
=
\mathrm{blkdiag}\!\left(
\mathbf S^{(h_1)},
\mathbf S^{(h_2)},
\dots,
\mathbf S^{(h_{|\mathcal H|})}
\right)\in\mathbb{C}^{|\mathcal{H}|N \times |\mathcal{H}|N}$, where $\mathbf{S}^{(h)}\in\mathbb{C}^{N \times N}$ characterizes the scattering within the subsystem at frequency $f_h$ and $N=N_\mathrm{A}+N_\mathrm{S}$. In contrast, the subsystem comprising the tunable components is time-varying such that its time-Floquet scattering matrix 
\begin{equation}
\widetilde{\mathbf{\Phi}}=
\begin{bmatrix}
\mathbf\Phi^{(h_1,h_1)} & \cdots & \mathbf\Phi^{(h_1,h_{|\mathcal H|})}\\
\vdots & \ddots & \vdots\\
\mathbf\Phi^{(h_{|\mathcal H|},h_1)} & \cdots & \mathbf\Phi^{(h_{|\mathcal H|},h_{|\mathcal H|})}
\end{bmatrix}\in\mathbb{C}^{|\mathcal{H}|N_\mathrm{S} \times |\mathcal{H}|N_\mathrm{S}}
\end{equation}
is generally \textit{neither block-diagonal nor symmetric}. 
For our independent element-wise modulation, each block $\mathbf\Phi^{(h_n,h_m)}$ is diagonal, with the diagonal entries given by the corresponding Fourier coefficients of the time-varying load reflection coefficients:
\begin{equation}
\mathbf\Phi^{(h_n,h_m)}
=
\operatorname{diag}\!\left(
 \phi_1^{(h_n,h_m)},
 \dots,
 \phi_{N_\mathrm{S}}^{(h_n,h_m)}
 \right),
\end{equation}
where
\begin{equation}
\phi_i^{(h_n,h_m)}
\!=\!
\frac{\mathrm{e}^{\jmath\theta_i^{(h_n,h_m)}}}{T_\mathrm{m}}
\sum_{q=1}^{Q}
\int_{(q-1)\Delta}^{q\Delta}
 \!r_i^{(q,h_m)}
 \mathrm{e}^{-\jmath 2\pi (h_n-h_m)t/T_\mathrm{m}}
\,\mathrm{d}t.
\label{eq_3}
\end{equation}
Here, \(r_i^{(q,h_m)}=\rho_{C_{i,q}}^{(h_m)}\) is the reflection coefficient of
the \(i\)th load for an incident wave at \(f_{h_m}\) during the \(q\)th slot,
where \(\rho_p^{(h)}\in\mathbb{C}\) with \(p\in\{1,\dots,P\}\) denotes the
reflection coefficient of the \(p\)th load state at frequency \(f_h\).
Moreover, \(\theta_i^{(h_n,h_m)}= - 2\pi (h_n-h_m)\tau_i/T_\mathrm{m}\).
The phase factor \(\mathrm{e}^{\jmath\theta_i^{(h_n,h_m)}}\) follows from the Fourier
time-shift property and accounts for the control delay $\tau_i$ of the \(i\)th
element.

Finally, given $\widetilde{\mathbf{S}}$ and $\widetilde{\mathbf{\Phi}}$, standard MNT arguments yield
\begin{equation}
\widetilde{\mathbf H}
=
\widetilde{\mathbf S}_{\widetilde{\mathcal{R}}\widetilde{\mathcal{T}}}
+
\widetilde{\mathbf S}_{\widetilde{\mathcal{R}}\widetilde{\mathcal{S}}}
\bigl(
\mathbf I_{|\mathcal H|N_\mathrm{S}}
-
\widetilde{\mathbf\Phi}\,
\widetilde{\mathbf S}_{\widetilde{\mathcal{S}}\widetilde{\mathcal{S}}}
\bigr)^{-1}
\widetilde{\mathbf\Phi}\,
\widetilde{\mathbf S}_{\widetilde{\mathcal{S}}\widetilde{\mathcal{T}}}.
\label{eq:Htilde}
\end{equation}
Here, \(\mathcal T\), \(\mathcal R\), and \(\mathcal S\) denote the port-index
sets associated with transmitting antennas, receiving antennas, and RIS elements,
respectively, while \(\widetilde{\mathcal T}\), \(\widetilde{\mathcal R}\), and
\(\widetilde{\mathcal S}\) collect the corresponding indices over all retained
harmonics.
For the MC-unaware benchmark, we set $\widetilde{\mathbf{S}}_{\widetilde{\mathcal{S}}\widetilde{\mathcal{S}}}=\mathbf{0}$.

\section{Estimation of Proxy TF-MNT Parameters}

\subsection{Estimation of Proxy MNT Parameters for Conventional RIS}

For a conventional RIS, $Q=1$ such that $\mathbf{C}$ collapses to a vector $\mathbf{c}\in\{1,\dots,P\}^{N_\mathrm{S}}$ and there is no frequency mixing. Thus, the proxy MNT parameters of a conventional RIS can be estimated separately for each considered frequency. For concreteness, we consider the fundamental frequency in the following. The parameters to be estimated include $\mathbf{H}_\mathrm{d}\triangleq\mathbf{S}_\mathcal{RT}^{(0)}$, $\mathbf{A}\triangleq\mathbf{S}_\mathcal{RS}^{(0)}$, $\mathbf{\Gamma}\triangleq\mathbf{S}_\mathcal{SS}^{(0)}$, $\mathbf{B}\triangleq\mathbf{S}_\mathcal{ST}^{(0)}$, and $\{\rho_p^{(0)}\}_{p=1}^P$. As mentioned, various techniques have been proposed to estimate a set of proxy MNT parameters based on measurements of $\mathbf{H}^{(0)}(\mathbf{c})$ for a series of known $\mathbf{c}$~\cite{sol2024experimentally,del2025experimental,del2026ambiguity,del2026reduced}.
A proxy parameter set $\hat{\boldsymbol\theta}^{(0)} \triangleq (\hat{\mathbf H}_\mathrm{d},\hat{\mathbf A},\hat{\mathbf \Gamma},\hat{\mathbf B},\{\hat{\rho}_p^{(0)}\}_{p=1}^P)$ is operationally equivalent to the true parameter set $\boldsymbol\theta^{(0)} \triangleq (\mathbf H_\mathrm{d},\mathbf A,\mathbf \Gamma,\mathbf B,\{{\rho}_p^{(0)}\}_{p=1}^P)$ if it yields the same end-to-end mapping: $\mathbf H^{(0)}(\mathbf c;\hat{\boldsymbol\theta}^{(0)}) = \mathbf H^{(0)}(\mathbf c;\boldsymbol\theta^{(0)})$ for all admissible choices of $\mathbf{c}$. The following three ambiguity classes generate valid proxy parameterizations [Sec.~III,~\cite{salmi2026electromagnetically}].

\textit{Diagonal-similarity (DS) gauge:} 
\begin{equation}
\label{eq:gauge_DS_mimo}
\begin{alignedat}{2}
\hat{\mathbf H}_\mathrm{d}
&= \mathbf H_\mathrm{d},\quad
&&\hat{\mathbf A} = \mathbf A\mathbf D^{-1},\quad
  \hat{\mathbf B} = \mathbf D\mathbf B,\\
\hat{\mathbf \Gamma}
&= \mathbf D\mathbf \Gamma \mathbf D^{-1},\quad
&&\hat\rho_p^{(0)} = \rho_p^{(0)} \ \forall\,p\in\{1,\dots,P\},
\end{alignedat}
\end{equation}
where $\mathbf{D}=\mathrm{diag}(\mathbf{d})$ and $\mathbf{d}\in \mathbb{C}^{ N_\mathrm{S}}$ is an arbitrary vector for which $\mathbf{D}$ is invertible.

\textit{Complex-scaling (CS) gauge:}
\begin{equation}
\label{eq:gauge_CS_mimo}
\begin{alignedat}{2}
\hat{\mathbf H}_\mathrm{d}
&= \mathbf H_\mathrm{d},\quad
&&\hat{\mathbf A} = \frac{1}{\gamma}\mathbf A,\quad
  \hat{\mathbf B} = \mathbf B,\\
\hat{\mathbf \Gamma}
&= \frac{1}{\gamma}\mathbf \Gamma,\quad
&&\hat\rho_p^{(0)} = \gamma\,\rho_p^{(0)} \ \forall\,p\in\{1,\dots,P\},
\end{alignedat}
\end{equation}
where \(\gamma\in\mathbb C\setminus\{0\}\) is an arbitrary non-zero scalar.

\textit{M\"obius (M\"O) gauge:}
\begin{equation}
\label{eq:gauge_MO_mimo}
\begin{alignedat}{2}
\hat{\mathbf H}_\mathrm{d}
&= \mathbf H_\mathrm{d} + \mu\,\mathbf A\mathbf F\mathbf B,\quad
&&\hat{\mathbf A} = k\,\mathbf A\mathbf F,\quad
  \hat{\mathbf B} = k\,\mathbf F\mathbf B,\\
\hat{\mathbf \Gamma}
&= (\mathbf \Gamma - \mu^*\mathbf I_{N_{\mathrm S}})\mathbf F,\quad
&&\hat\rho_p^{(0)} = \mathcal M_{\mu}(\rho_p^{(0)}) \ \forall\,p\in\{1,\dots,P\},
\end{alignedat}
\end{equation}
where $\mathcal M_{\mu}(\rho)\triangleq \frac{\rho-\mu}{1-\mu^*\rho}$, 
$\mathbf F\triangleq (\mathbf I_{N_{\mathrm S}}-\mu\mathbf \Gamma)^{-1}$, 
and $k\triangleq \sqrt{1-|\mu|^2}$, and \(\mu\in\mathbb C\) is an arbitrary scalar subject to the constraints that all inverses required to evaluate \eqref{eq:gauge_MO_mimo} exist, that $k\neq0$, and that $1-\mu^*\rho_p^{(0)}\neq 0$ for all \(p\in\{1,\dots,P\}\).

For the MC-unaware model with
\(\mathbf{\Gamma}=\mathbf 0\), the DS and CS gauges  remain admissible (DS and CS map \(\mathbf{\Gamma}=\mathbf 0\) to
\(\hat{\mathbf{\Gamma}}=\mathbf 0\)), while the M\"obius gauge is not 
admissible (M\"O maps $\mathbf{\Gamma}$ to 
\(\hat{\mathbf{\Gamma}}=-\mu^*\mathbf I_{N_{\mathrm S}}\)). Therefore, in the MC-unaware case, the M\"obius gauge must be replaced by an \textit{affine-shift (AF) gauge}:
\begin{equation}
\label{eq:gauge_AF_mimo}
\begin{alignedat}{2}
\hat{\mathbf H}_\mathrm{d}
&= \mathbf H_\mathrm{d}-\eta\,\mathbf A\mathbf B,\quad
&&\hat{\mathbf A} = \mathbf A,\quad
  \hat{\mathbf B} = \mathbf B,\\
\hat{\mathbf \Gamma}
&= \mathbf 0,\quad
&&\hat\rho_p^{(0)} = \rho_p^{(0)}+\eta \ \forall\,p\in\{1,\dots,P\},
\end{alignedat}
\end{equation}
where \(\eta\in\mathbb C\) is an arbitrary scalar.

\subsection{Estimation of Proxy TF-MNT Parameters for TF-RIS}

\textit{Step 1:} For each considered harmonic in turn, we independently estimate a set of proxy MNT parameters using the technique presented in~\cite{del2025experimental}. We thus operate the TF-RIS as conventional RIS in Step 1 (or, equivalently, a TF-RIS with $Q=1$). Step 1 can be fully parallelized across the considered harmonics, both in terms of measurements and processing.

\textit{Step 2:} Having estimated the conventional proxy MNT parameters independently at each retained harmonic, we now align their ambiguity gauges using measurements with dynamic TF-RIS modulation patterns (i.e., $Q > 1$). Depending on the measurement mode, we measure $\mathcal{P}(\widetilde{\mathbf{H}}(\mathbf{C}_k))$ for \(K\) random TF-RIS modulation patterns \(\{\mathbf C_k\}_{k=1}^{K}\). We consider three measurement modes:

\textit{M1:} Only the complex-valued fundamental-to-fundamental block of the end-to-end channel is measured: \(\mathcal P_1(\widetilde{\mathbf H})\) extracts from \(\widetilde{\mathbf H}\) the block mapping inputs at \(f_0\) to outputs at \(f_0\). This is sufficient when the TF-RIS is used to create additional degrees of freedom at the fundamental frequency, without exploiting the resulting harmonic signals. Experimentally, M1 can rely on a vector network analyzer, with the intermediate-frequency bandwidth chosen much smaller than \(f_\mathrm{m}\).

\textit{M2:} Only the magnitudes of the multi-frequency end-to-end channel matrix are measured: \(\mathcal P_2(\widetilde{\mathbf H})=|\widetilde{\mathbf H}|\), where the modulus is applied elementwise. This mode is relevant for applications such as harmonic beam steering, where the objective depends on harmonic powers rather than on coherent multi-harmonic phase relations. Experimentally, M2 can be realized with a signal generator and a spectrum analyzer. For a MIMO measurement, the transmit excitation is switched across the transmit ports and retained input harmonics, while the spectrum analyzer is switched across the receive ports and spans the retained output harmonics.

\textit{M3:} The complex-valued multi-frequency end-to-end channel is measured: \(\mathcal P_3(\widetilde{\mathbf H})=\widetilde{\mathbf H}\). This mode is required for coherent multi-harmonic processing or phase-sensitive harmonic combining. Experimentally, M3 can be realized with phase-coherent multi-harmonic acquisition, for example using a signal generator and a phase-referenced high-sampling-rate oscilloscope. For a MIMO measurement, the transmit excitation is switched across the transmit ports and retained input harmonics, while the receiver is switched across the receive ports, or records them in parallel, and extracts the complex amplitudes at the retained output harmonics.

Let \(\hat{\boldsymbol\theta}^{(h)}\) be the proxy parameter set estimated at harmonic \(h\) in Step 1. We now apply harmonic-dependent ambiguity transformations parametrized by \(\boldsymbol\phi^{(h)}=(\mathbf d^{(h)},\gamma^{(h)},\mu^{(h)})\in\mathbb C^{N_\mathrm{S}+2}\), where \(\mathbf d^{(h)}\), \(\gamma^{(h)}\), and \(\mu^{(h)}\) correspond to the DS, CS, and M\"O gauges, respectively.\footnote{In the MC-unaware case, we replace $\mu^{(h)}$ from M\"O by $\eta^{(h)}$ from AF.} Let \( g\) denote the composition of the admissible transformations, applied in the order DS--CS--M\"O. We identify an optimized set of gauge parameters $\{\boldsymbol\phi_\star^{(h)}\}_{h\in\mathcal H}$ by solving
\begin{equation}
\label{eq:cross_harmonic_alignment}
\begin{aligned}
\{\boldsymbol\phi_\star^{(h)}\}_{h\in\mathcal H}
&=
\arg\min_{\{\boldsymbol\phi^{(h)}\}_{h\in\mathcal H}}
\frac{1}{K}
\sum_{k=1}^{K}
\varepsilon_k\!\left(\{\boldsymbol\phi^{(h)}\}_{h\in\mathcal H}\right),\\
\varepsilon_k\!\left(\!\{\boldsymbol\phi^{(h)}\!\}_{h\in\mathcal H}\!\right) \!
&= \!
\left\|
\mathcal P\!\left(
\widetilde{\mathbf H}_{k}^{\mathrm{pred}}\!
\right)\!
-\!
\mathcal P\!\left(
\widetilde{\mathbf H}^{\mathrm{meas}}_k\!
\right)\!
\right\|_1
\big/
\left\|
\mathcal P\!\left(
\widetilde{\mathbf H}^{\mathrm{meas}}_k\!
\right)\!
\right\|_1.
\end{aligned}
\end{equation}
The minimization in \eqref{eq:cross_harmonic_alignment} is subject to the admissibility constraints of the DS, CS, and M\"O transformations stated above. The retained ambiguity-aligned proxy TF-MNT parameters are defined as $\boldsymbol\theta_\star^{(h)} =  g\!\left( \hat{\boldsymbol\theta}^{(h)}; \boldsymbol\phi_\star^{(h)} \right)$ for $ h\in\mathcal H$.
The unknown control delays \(\{\tau_i\}_{i=1}^{N_\mathrm{S}}\) need not be estimated separately because
\(\mathrm{e}^{-\jmath2\pi(h_n-h_m)\tau_i/T_\mathrm{m}}\) factorizes into output- and input-harmonic diagonal phase factors, which are absorbed by the harmonic-dependent DS gauges.
For the MC-unaware benchmark model, we use the same cross-harmonic alignment procedure, but we change the admissible gauge set from DS--CS--M\"O to DS--CS--AF, applied in the same order.
A cross-harmonic ambiguity alignment valid for M3 is also valid for M1 and M2, but not conversely in general.

We solve \eqref{eq:cross_harmonic_alignment} in TensorFlow using a first-order adaptive gradient optimizer (full-batch Adam). We initialize the real and imaginary parts of \(d_i^{(h)}\) and \(\gamma^{(h)}\) from truncated normal distributions centered at \(1\) and \(0\), respectively, with standard deviation \(2\times10^{-2}\); \(\mu^{(h)}\) or \(\eta^{(h)}\) is initialized analogously, but with both parts centered at \(0\). 
We decrease the step size from \(10^{-3}\) to \(10^{-5}\) over the course of the utilized 250 iterations.

\section{Full-Wave Numerical Validation and \\Performance Analysis}

\begin{figure}
    \centering
    \includegraphics[width=\columnwidth]{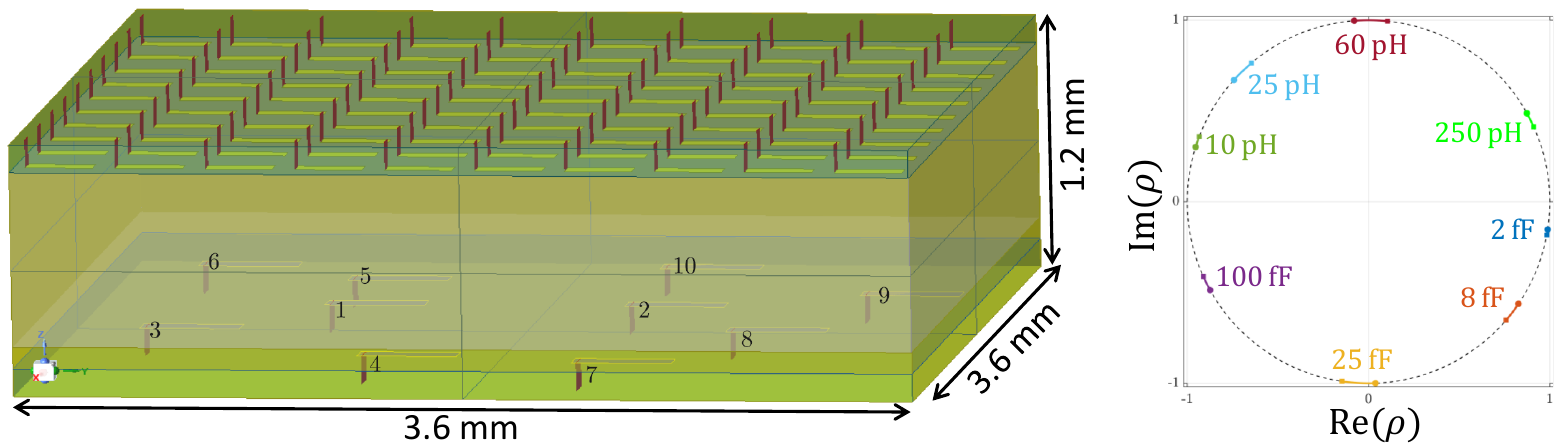}
    \caption{Left: Wireless network-on-chip setup from~\cite{hammami2026expressivity} with 10 antennas (bottom) and 100 RIS elements (top). Right: Dispersion of eight load states from 122.5~GHz to 147.5~GHz.}
    \label{Fig2}
\end{figure}

\textit{Setup:} 
We consider the D-band RIS-parametrized wireless network-on-chip setup from~\cite{hammami2026expressivity} shown in Fig.~\ref{Fig2}, comprising 10 antennas and 100 RIS elements. The corresponding 110-port scattering matrix was obtained via full-wave numerical simulation in ANSYS HFSS at all considered harmonics~\cite{hammami2026expressivity}. Throughout this Letter, we only consider scenarios involving eight antennas and ten RIS elements, assuming that the ports of the remaining unused antennas and RIS elements are terminated in matched loads. We consider 3-bit programmable RIS elements with dispersive loads, as shown in Fig.~\ref{Fig2}. We use $f_0=135$~GHz, $f_\mathrm{m}=125$~MHz, and we define the ground-truth harmonic end-to-end channels with $|\mathcal H|=201$. We control the signal-to-noise ratio (SNR) by adding white circular complex Gaussian noise to the complex-valued harmonic end-to-end channel matrices before applying the measurement operator \(\mathcal P\).

\textit{Proxy TF-MNT Parameter Estimation:}
To quantify the accuracy with which a given model can predict the end-to-end
harmonic channels, following~\cite{del2025experimental,del2026ambiguity}, we compute
\begin{equation}
\zeta_\mathcal{P}
=
\frac{
\operatorname{SD}\!\left(
\operatorname{vec}\!\left(
\left\{\mathcal{P}(\widetilde{\mathbf{H}}_{\ell}^{\mathrm{true}})\right\}_{\ell=1}^{100}
\right)
\right)
}{
\operatorname{SD}\!\left(
\operatorname{vec}\!\left(
\left\{
\mathcal{P}(\widetilde{\mathbf{H}}_{\ell}^{\mathrm{true}})
-
\mathcal{P}(\widetilde{\mathbf{H}}_{\ell}^{\mathrm{pred}})
\right\}_{\ell=1}^{100}
\right)
\right)
},
\label{eq:defineZeta}
\end{equation}
where \(\operatorname{SD}(\cdot)\) is evaluated over all entries after stacking
the selected observables from 100 unseen random TF-RIS modulation patterns.
The metric is analogous to an SNR, with model errors treated as effective noise.

\begin{figure}
    \centering
    \includegraphics[width=\columnwidth]{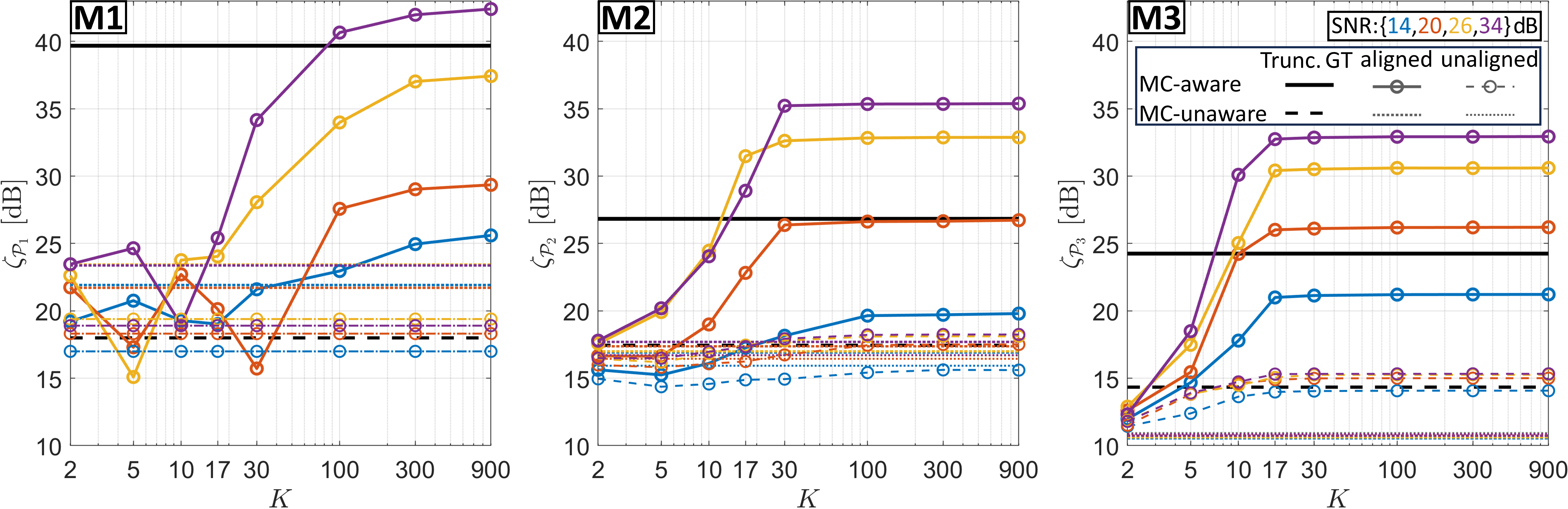}
    \caption{Dependence of $\zeta_\mathcal{P}$ on $K$, SNR, and measurement mode. All displayed results are for $N_\mathrm{T}=N_\mathrm{R}=4$, $N_\mathrm{S}=10$, $Q=3$, and $|\mathcal H|=11$. Curves are displayed for the ground-truth (GT) model truncated to $|\mathcal H|=11$, the aligned proxy TF-MNT model, and the unaligned proxy TF-MNT model. The MC-aware and MC-unaware cases are compared in all scenarios.}
    \label{Fig3}
\end{figure}

We begin by examining in Fig.~\ref{Fig3} the dependence of $\zeta_\mathcal{P}$ on $K$ and the SNR, across the three considered measurement modes. For concreteness, we fix $N_\mathrm{T}=N_\mathrm{R}=4$, $N_\mathrm{S}=10$, $Q=3$, and $|\mathcal H|=11$. As a benchmark, we use the ground-truth TF-MNT model truncated to \(|\mathcal H|=11\). We furthermore show the accuracy without cross-harmonic ambiguity alignment. In addition, we include the counterparts of all these results for the MC-unaware model. 
\textit{First}, we observe that the truncated ground-truth models retaining \(|\mathcal H|=11\) harmonics only reach finite accuracies due to the truncation of higher harmonics; this truncation is most significant for M3, where the truncated ground-truth model reaches only 24.2~dB, compared with 26.8~dB for M2 and 39.7~dB for M1. 
\textit{Second}, we see that the cross-harmonic ambiguity-aligned TF-MNT model can significantly exceed these accuracies, reaching up to 32.9~dB for M3, i.e., 8.7~dB above the truncated ground-truth model. This is possible because the proxy parameters can absorb part of the truncation error caused by neglected higher-order harmonics. Thus, when the number of retained harmonics is limited to reduce the computational complexity of TF-MNT, ambiguity-aligned proxy parameters can be preferable to naively truncated ground-truth parameters. The required SNR level for the aligned proxy TF-MNT model to outperform the truncated ground-truth model is highest for M1 and lowest for M3.
\textit{Third}, we see that the number of required measurements beyond which the curves saturate is roughly $K=17$ for M3, roughly $K=30$ for M2, and roughly $K=900$ for M1. This makes sense, because the amount of information acquired per measurement in M3 is twice that acquired in M2 and $11^2=121$ times that acquired in M1. 
\textit{Fourth}, we notice that MC unawareness leads to roughly 10~dB lower accuracy with the truncated ground-truth model for M2 and M3, and up to roughly 18~dB lower accuracy with the cross-harmonic ambiguity-aligned proxy TF-MNT model. 
\textit{Fifth}, we see that the cross-harmonic ambiguity alignment is pivotal; without it, the accuracy drops by up to roughly 22~dB in M3.

\begin{figure}
    \centering
    \includegraphics[width=\columnwidth]{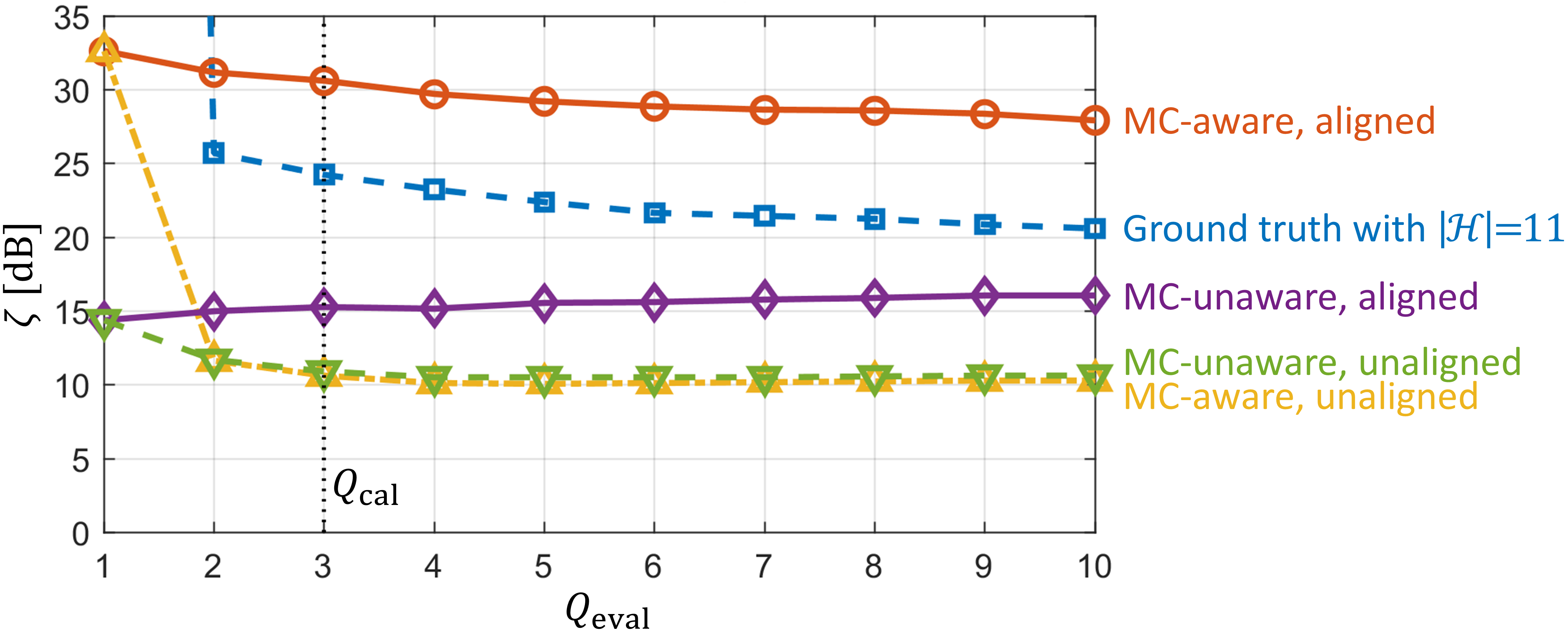}
    \caption{Dependence of $\zeta_\mathcal{P}$ on $Q_\mathrm{eval}$ for proxy parameters that were ambiguity-aligned at $Q_{\rm cal}=3$. All displayed results are for $K=900$, SNR~$=26$~dB, M3.}    
    \label{Fig4}
\end{figure}

Next, we examine in Fig.~\ref{Fig4} whether proxy TF-MNT parameters estimated for $Q_\mathrm{cal}=3$ are also valid for $Q_\mathrm{eval}\neq 3$. Curves for $Q_\mathrm{cal}=5$ were nearly identical and are thus omitted for clarity. It is apparent that MC-aware proxy parameters with cross-harmonic ambiguity alignment work well for all considered values of $Q_\mathrm{eval}$, with a slight decline from 32.6~dB at $Q=1$ to 27.9~dB at $Q=10$. A similar decline is also visible for the truncated ground-truth parameter model (except that for $Q=1$ its accuracy tends to infinity since there is no harmonic truncation effect). Without ambiguity alignment, the accuracy is roughly 20~dB lower for $Q>1$. Interestingly, the accuracy of the MC-unaware model with cross-harmonic ambiguity alignment slightly improves from 14.4~dB at $Q=1$ to 16.1~dB at $Q=10$; however, it remains at least 11.8~dB below its MC-aware counterpart. Without cross-harmonic ambiguity alignment, MC awareness makes no notable difference for $Q>1$, with $\zeta_\mathcal{P}$ in the vicinity of 10~dB.

\textit{TF-RIS Performance Evaluation:} 
To assess the impact of cross-harmonic ambiguity alignment and MC awareness on performance, we consider the prototypical problem of maximizing the single-input single-output (SISO) fundamental-to-first-harmonic channel gain using proxy parameters estimated from $K=900$ measurements with SNR~\(=26\)~dB. We use the problem statement [(P2),~\cite{kuznetsov2026mutual}] and the coordinate-ascent algorithm [Algorithm~1,~\cite{kuznetsov2026mutual}]. 
Our results are summarized in Table~\ref{table1} for M2; the results for M3 are essentially identical. 
The best performance is achieved and accurately predicted by our proxy TF-MNT model: $-21.0$~dB with $Q=3$ and $-19.8$~dB with $Q=10$. The performance is minimally better than with the truncated ground-truth TF-MNT model, 5.33~dB better than without cross-harmonic ambiguity alignment (for $Q=3$). Relative to the MC-aware ambiguity-aligned case, the performance drops by 2.5~dB without MC awareness, by 5.3~dB without ambiguity alignment, and by 11.9~dB when neither is used. Altogether, misaligned cross-harmonic ambiguities appear to harm the performance more than MC unawareness.

\begin{table}
\centering
\caption{True \textcolor{gray}{[pred.]} fundamental-to-first-harmonic SISO channel gain with optimized TF-RIS modulation pattern in dB.}
\label{table1}
\setlength{\tabcolsep}{2.1pt}
\renewcommand{\arraystretch}{1.05}
\begin{tabular}{llcc}
\toprule
Model &  & \(Q_\mathrm{eval}=3\) & \(Q_\mathrm{eval}=10\) \\
\midrule
\multirow{3}{*}{MC-aware}
& Trunc. GT & $-21.06$ \textcolor{gray}{\scriptsize[$-21.01$]} & $-19.79$ \textcolor{gray}{\scriptsize[$-19.69$]} \\
& Aligned   & $\mathbf{-21.00}$ \textcolor{gray}{\scriptsize[$-20.98$]} & $\mathbf{-19.80}$ \textcolor{gray}{\scriptsize[$-19.69$]} \\
& Unaligned & $-26.33$ \textcolor{gray}{\scriptsize[$-21.76$]} & $-24.73$ \textcolor{gray}{\scriptsize[$-20.44$]} \\
\midrule
\multirow{3}{*}{MC-unaware}
& Trunc. GT & $-25.35$ \textcolor{gray}{\scriptsize[$-24.75$]} & $-23.65$ \textcolor{gray}{\scriptsize[$-23.36$]} \\
& Aligned   & $-23.47$ \textcolor{gray}{\scriptsize[$-20.41$]} & $-22.80$ \textcolor{gray}{\scriptsize[$-19.32$]} \\
& Unaligned & $-32.90$ \textcolor{gray}{\scriptsize[$-20.84$]} & $-31.96$ \textcolor{gray}{\scriptsize[$-19.66$]} \\
\bottomrule
\end{tabular}
\end{table}

\section{Conclusion}

To summarize, we presented a segmented technique for estimating ambiguity-aligned proxy TF-MNT parameters of TF-RIS-parametrized radio environments from end-to-end measurements. Our method combines independent per-harmonic proxy MNT estimation with cross-harmonic ambiguity alignment. We corroborated the importance of this ambiguity alignment in terms of model accuracy and performance based on full-wave numerical results.

\vspace{-2mm}
\section*{Acknowledgment}
The author acknowledges stimulating discussions with A.~D.~Kuznetsov and V.~Viikari.
\vspace{-2mm}

\bibliographystyle{IEEEtran}
%\bibliography{references}

% Generated by IEEEtran.bst, version: 1.14 (2015/08/26)
\providecommand{\noopsort}[1]{}\providecommand{\singleletter}[1]{#1}%

\end{document}